\documentclass{article}

\usepackage{amssymb}

\usepackage{arxiv}

\usepackage[utf8]{inputenc} % allow utf-8 input
\usepackage[T1]{fontenc}    % use 8-bit T1 fonts
\usepackage{hyperref}       % hyperlinks
\usepackage{url}            % simple URL typesetting
\usepackage{booktabs}       % professional-quality tables
\usepackage{amsfonts}       % blackboard math symbols
\usepackage{nicefrac}       % compact symbols for 1/2, etc.
\usepackage{microtype}      % microtypography
\usepackage{lipsum}		% Can be removed after putting your text content
\usepackage{graphicx}
\usepackage[sort,square,numbers]{natbib}
\usepackage{doi}

\begin{document}

%\tytul{Additional look into GAN-based augmentation for deep learning COVID-19 image classification\ldots} % do not capitalize all the words except where needed
%\autor{O.\ Fedoruk, K.\ Klimaszewski, A. Ogonowski, M.\ Kruk} % for page headings, initials of given names and full family names, separated by commas
\title{Additional Look into GAN-based Augmentation for Deep Learning COVID-19 Image Classification}

% capitalize all the words, according to the typical rules for titles
\author{Oleksandr Fedoruk$^1$, Konrad Klimaszewski$^1$, Aleksander Ogonowski$^1$, Michał Kruk$^2$\\ % full names, with full given names and middle names (if necessary; middle names can
% be represented by initials), separated with commas. See the example below.
{\footnotesize\sl $^1$Department of Complex Systems, National Centre for Nuclear Research, Otwock-Świerk, Poland}\\
{\footnotesize\sl $^2$Institute of Information Technology, Warsaw University of Life Sciences, Warsaw, Poland}
% affiliation consists of: the name of the Institution, City, Country;
% you can add www or corresponding author's e-mail using \href{}{} from the hyperref package
}
% if there is more than one Author, and more than one affiliation, please use upper indexes
% to relate Authors to Affiliations, for example:
%\author{Givenname M. Familyname$^1$, Givenname M. Familyname$^2$\\
%  {\footnotesize\sl $^1$Department, University, City, Country}\\ % take each affiliation in separate braces {.}
%  {\footnotesize\sl $^2$Department, University, City, Country}
%  {\footnotesize\sl \href{mailto:corresponding.author[at]mail.domain}{corresponding.author@mail.domain}}
%}

\maketitle

\pagestyle{myheadings}

\noindent
{\footnotesize{\bf Abstract.}\ \ The availability of training data is one of the main limitations in deep learning applications for medical imaging. Data augmentation is a popular approach to overcome this problem. Classical augmentation is based on modification (rotations, shears, brightness and contrast changes, etc.) of the images from the original dataset. Another possible approach is a Machine Learning based augmentation, in particular usage of Generative Adversarial Networks (GAN). In this case, GANs generate images similar to the original dataset so that the overall training data amount is bigger, which leads to better performance of trained networks. A GAN model consists of two networks, a generator and a discriminator. These two networks are interconnected in a feedback loop which creates a competitive environment. The generator attempts to synthesize realistic images. Meanwhile, the discriminator attempts to distinguish between real and synthetic images as precisely as possible. This process results in the generator continually improving its ability to create realistic images. 
This work is a continuation of the previous research where we trained StyleGAN2-ADA by Nvidia on the limited COVID-19 chest X-ray image dataset. In this paper, we study the dependence of the GAN-based augmentation performance on dataset size with a focus on small samples. Two datasets are considered, one with 1000 images per class (4000 images in total) and the second with 500 images per class (2000 images in total). 
We train StyleGAN2-ADA with both sets and then, after validating the quality of generated images, we use trained GANs as one of the augmentations approaches in multi-class classification problems. We compare the quality of the GAN-based augmentation approach to two different approaches (classical augmentation and no augmentation at all) by employing transfer learning-based classification of COVID-19 chest X-ray images. The results are quantified using different classification quality metrics and compared to the results from the previous article and literature. The GAN-based augmentation approach is found to be comparable with classical augmentation in the case of medium and large datasets but underperforms in the case of smaller datasets. The correlation between the size of the original dataset and the quality of classification is visible independently from the augmentation approach.  
\par}

\medskip{\centerline{\emph{Submitted to Machine Graphics \& Vision}}}

\medskip{\footnotesize{\bf Key words:}\ \ computer vision, deep learning, image classification, generative adversarial networks, medical imaging \par}

\section{Introduction} 
\label{sec:intro}

Computer vision techniques are used in different medical applications for various purposes. They accelerate decision-making while diagnosing patients and support medical personnel on a daily basis. As available algorithms and solutions advance, the problem of medical data accessibility is becoming a bottleneck for new researchers and breakthroughs~\cite{collectionOfMedicalImageDatasets}. Almost all modern algorithms are data-driven and require a lot of data samples to perform well ~\cite{whang2022data}. However, the process of gathering medical data is not easy and often blocked by the high costs of procedures required to obtain the data, patients' personal data access limitations (as GDPR in the European Union or CCPA in the United States of America), rare diseases for which there is just not much data at all. 

To overcome that problem, researchers use data augmentation techniques. The main idea is to train models on several modified copies of original data. This reduces overfitting and allows to achieve better training results with less data ~\cite{dataAugmentation}. In the graphical data domain, the classical augmentation pipeline includes transformations such as rotating, scaling, changing the brightness of the image, etc. With the rapid development of Generative Adversarial Networks (GAN) \cite{goodfellow2014generative} it is possible to generate images similar to ones from the given dataset so it is possible to apply GANs as an augmentation pipeline. 

In our original experiment, we showed, on a dataset that contains 2000 images per class, that the GAN-based augmentation approach is comparable to but not outperforming classical augmentation~\cite{FirstPaper}. In this paper, we want to move forward and verify if the same is true for even smaller datasets.

The manuscript is organized as follows: in the \textit{Introduction} the research problem is described; the \textit{Methods} section describes the \textit{Motivation and methodology} and the \textit{Dataset} including its preprocessing followed by \textit{Image comparison metrics} used to assess the quality of the \textit{GAN-based augmentation} as well as the \textit{Classical augmentation}, both described in detail; the section concludes with a description of the approach used for the \textit{Classification} of images.
In the \textit{Results} section dependence of the classifier output on the sample size is described.
The article ends with a \textit{Conclusion}.

\section{Methods}
% This template for Machine Graphics \&{} Vision was compiled on \today. The most up-to-date version can be found in~\cite{bib:MGV-template}.

\subsection{Motivation and methodology}

In this paper, we test the hypothesis that data augmentation plays a more crucial role in the deep learning process with small datasets than it does with large ones. Also, we try to verify that, in the case of medical imaging, data augmentation based on GAN-generated images could result in bigger data diversity and thus improve deep learning results in comparison to the classical augmentation approach.

We compare 3 data-augmentation approaches with two datasets - the first dataset contains 1000 images per 4 classes and the second contains 500 images per 4 classes respectively. Further in the paper we call the dataset with 1000 images per class a "small" dataset. The dataset with 500 images per class is called "micro" respectively. To create the small dataset 1000 images of each class were randomly picked from the original dataset after the prepossessing. The micro dataset is a subset of the small dataset and was also created by picking 500 random images of each class. The datasets used are described in detail in the following section.

Data augmentation approaches being compared are: no augmentation at all, classical augmentation, and GAN-based augmentation. To estimate which approach is better we train a convolutional neural network on data augmented by each approach and evaluate based on different classification quality metrics. 

% In case of GAN-augmentation we've trained GAN with both datasets and expanded them with 1000 generated artificial images.
% In case of classical augmentation we've applied close to typical pipeline and also expanded original datasets with 1000 modified images each. 

\begin{figure}[tb] % avoid the location qualifier "h" unless strictly necessary
  \centering
  \includegraphics[width=0.7\textwidth]{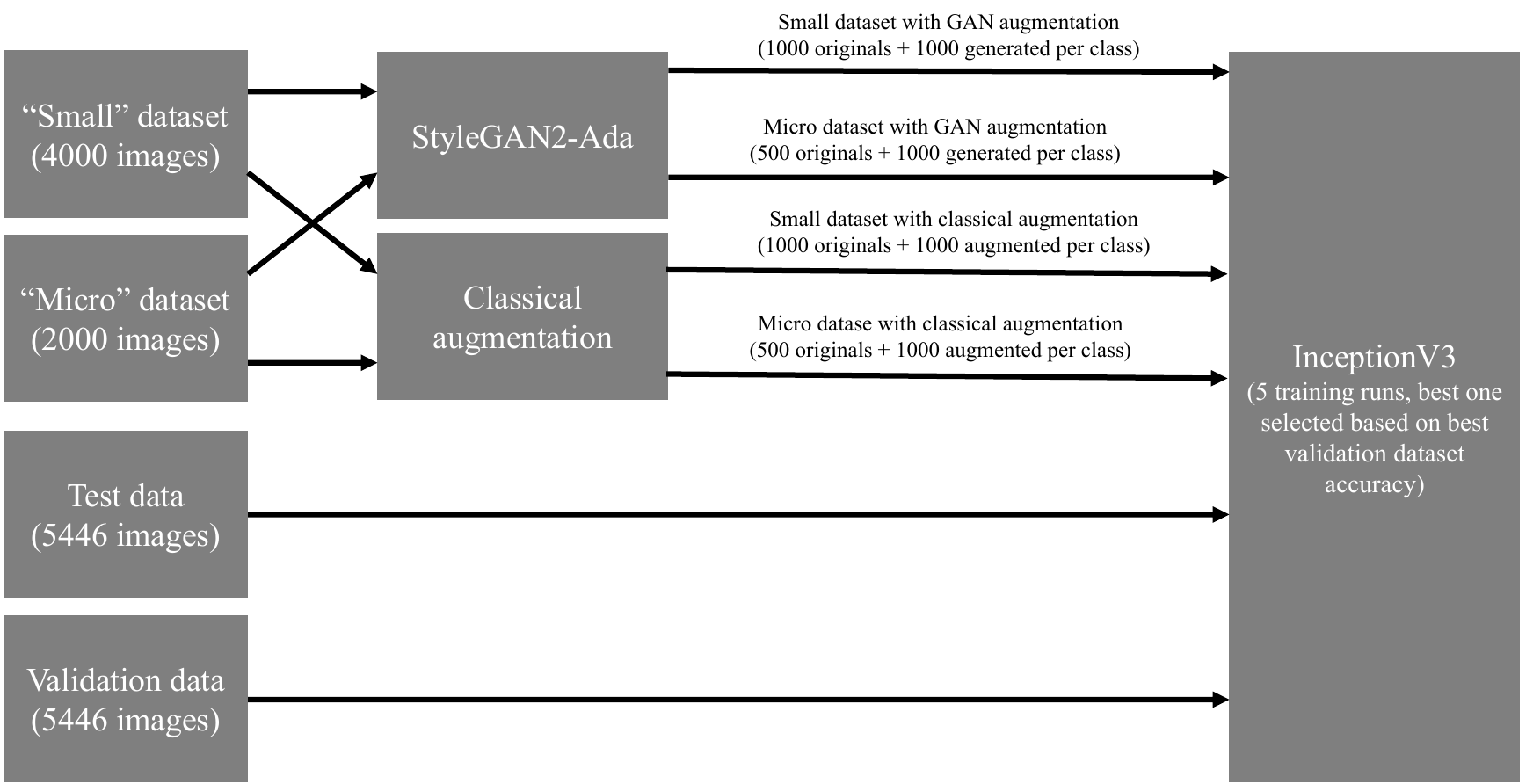} % recommended image formats are eps and pdf without lossless compression; please provide figures in which no artefacts are seen
  \caption{Visualisation of experiment steps and data flow.}
  \label{fig:experiment_scheme}
\end{figure}

\subsection{Dataset}

The dataset used in the research is the "COVID-19 Radiography Database" developed by a team of researchers from Qatar University, Doha, Qatar, and the University of Dhaka, Bangladesh along with their collaborators from Pakistan and Malaysia in collaboration with medical doctors. It is worth noticing that the dataset is being updated and the latest version of it contains way more images than it was when the original experiment was done. As this paper's goal is an additional investigation of the GAN augmentation technique, we continue to use the same version of the dataset that we used in the original experiment.

The dataset used in this paper, contains 3616 images of COVID-19-positive cases, 6012 images of lung opacity (non-COVID lung
infection), 1345 images of viral pneumonia, and 10192 images of healthy lungs. Each image is represented in PNG format with 256x256 dimensions. For each image, the dataset authors provided a corresponding lung
segmentation mask obtained using a dedicated U-Net model ~\cite{Rahman2021}. The dataset was split into 3 subsets: train, validation, and test. The small train subset contains 1000 (500 in the case of the micro dataset) images of each class and is used as a source for classical augmentation, training of the GAN, and for no augmentation approach. The rest of the images were split in half to form validation and test subsets. Also, we use the same test and validation subsets for both experiments with small and micro datasets. The validation subset is used as validation data in target classification CNN training. The test subset was used only as the final trained CNN benchmark which simulates new data from new patients to show "real-world" usage of the trained classification network. 

\begin{figure}[tb] % avoid the location qualifier "h" unless strictly necessary
  \centering
  % example of three subfigures; you can also use the macros for subfigures (their recent versions)
  % In this example the picture environment is used only to enable \framebox to make a frame around it,
  % only for this example. In typical cases simply use \includegraphics, no frame, as in the previous figure.
  \setlength{\unitlength}{1mm} % this is necessary only for the picture environment; delete if you don't use it
  {\bf a}\hspace{0.3em} \includegraphics[width=0.2\textwidth]{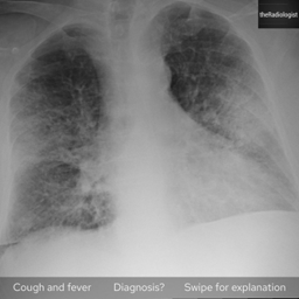} \hspace{1em}
  {\bf b}\hspace{0.3em} \includegraphics[width=0.2\textwidth]{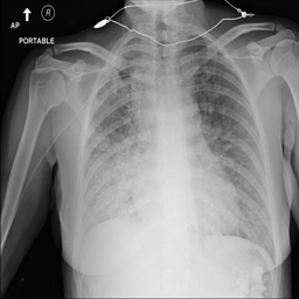} \\
  {\bf c}\hspace{0.3em} \includegraphics[width=0.2\textwidth]{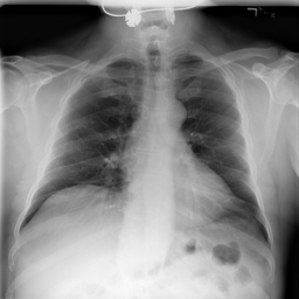} \hspace{1em} 
  {\bf e}\hspace{0.3em} \includegraphics[width=0.2\textwidth]{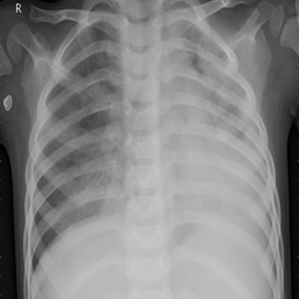} 
  \caption{
    Examples of unprocessed images from the original dataset. 
    ({\bf a})~COVID-19;
    ({\bf b})~Normal;
    ({\bf c})~Lung opacity;
    ({\bf e})~Viral pneumonia.
  }
  \label{fig:original_unprocessed_images}
\end{figure}

\begin{figure}[tb] % avoid the location qualifier "h" unless strictly necessary
  \centering
  % example of three subfigures; you can also use the macros for subfigures (their recent versions)
  % In this example the picture environment is used only to enable \framebox to make a frame around it,
  % only for this example. In typical cases simply use \includegraphics, no frame, as in the previous figure.
  \setlength{\unitlength}{1mm} % this is necessary only for the picture environment; delete if you don't use it
  {\bf a}\hspace{0.3em} \includegraphics[width=0.2\textwidth]{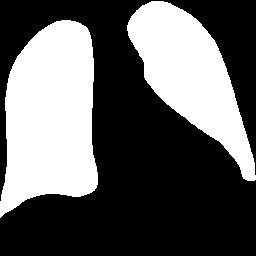} \hspace{1em}
  {\bf b}\hspace{0.3em} \includegraphics[width=0.2\textwidth]{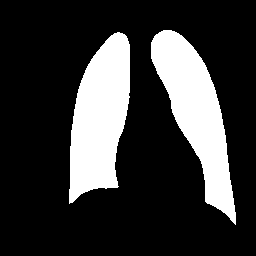} \\
  {\bf c}\hspace{0.3em} \includegraphics[width=0.2\textwidth]{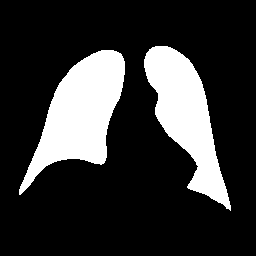} \hspace{1em} 
  {\bf e}\hspace{0.3em} \includegraphics[width=0.2\textwidth]{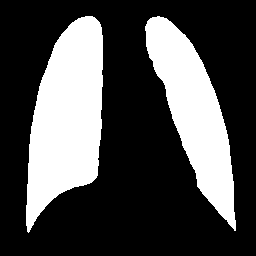} 
  \caption{
    Example binary masks from the original dataset for the images shown in ~\ref{fig:original_unprocessed_images}. 
  }
  \label{fig:masks_for_original_unprocessed_images}
\end{figure}

We've prepossessed all images from the original dataset with the following 3-step procedure:

\begin{enumerate}
    \item All images were cropped to the lung region according to the provided masks.
    \item Cropped images were manually reviewed and all images containing any text or graphical annotations/marks were removed.
    \item Remaining images were resized into 128x128 dimensions and converted to 1-channel (grayscale) to reduce the amount of data processed while training. The conversion was done by leaving only the first channel of the original images.
\end{enumerate}

After all the described steps, the final version of the dataset used in the experiment contained 3242 images of COVID-19 cases, 2982 images of lung opacity (non-COVID lung infection), 1264 images of viral pneumonia, and 7404 images of healthy lungs. As mentioned earlier 2 experimental cases have been considered in the research. The first experiment was conducted with the small dataset (1000 images per class in the training subset) and the second one with the micro dataset (500 images per class in the training subset). As the original dataset contained 4 classes, there are 4000 images in total in the train subset of the small dataset and 2000 images in the train subset of the micro dataset. Test and validation subsets remained the same for both cases and were prepared during small dataset preparation. It allowed us to be able to compare the final results for both experiments as the data those results were calculated on remained the same.

\begin{table}[tb] % avoid the location qualifier "h" unless strictly necessary
  \centering
  \caption{Amount of images per class in each subset for both small and micro datasets.}
  \label{tab:sometable}
  \begin{tabular}{lllll}
    \hline\hline
    & COVID-19 & Healthy & Lung Opacity & Viral Pneumonia \\ \hline
    Train subset 1000 & 1000 & 1000 & 1000 & 1000 \\
    Train subset 500 & 500 & 500 & 500 & 500 \\
    Validation subset & 1121 & 3202 & 991 & 132 \\
    Test subset & 1121 & 3202 & 991 & 132 \\
    \hline\hline
  \end{tabular}
\end{table}

\subsection{GAN-based augmentation}

In this research we continued to use StyleGAN2 with adaptive discriminator augmentation (ADA) mechanism by NVIDIA ~\cite{StyleGAN2ADA} as one of its features is the ability to be trained on relatively small datasets and support of class-conditional image generation. For each of the experiments, StyleGAN2-ADA was trained on the corresponding train subset. The training process was monitored, using the validation dataset, to prevent network overfitting. After the training, the epoch with the best Kernel Inception Distance (KID) score was picked as a source of future data generation. The target for $r_t$ ADA heuristic is set to 0.6, both generator and discriminator learning rates were set to 0.0025 while the batch size was set to 32. Following StyleGAN2 authors~\cite{StyleGAN2}, we use non-saturating logistic loss with $R_1$ regularisation with $gamma$ set to 1.024. Multiclass training was enabled so one network was able to generate images for all target classes after the training was done. All other parameters are set to default values provided by the NVIDIA implementation~\cite{StyleGAN2ADA}. The network was trained with a single NVIDIA Tesla K80 GPU and for each case, training took around 7 days.

\begin{figure}[tb]
  \centering
  
  \setlength{\unitlength}{1mm}
  {\bf a}\hspace{0.3em} \includegraphics[width=0.2\textwidth]{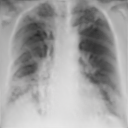} \hspace{1em}
  {\bf b}\hspace{0.3em} \includegraphics[width=0.2\textwidth]{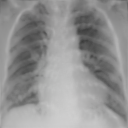} \hspace{1em}
  {\bf c}\hspace{0.3em} \includegraphics[width=0.2\textwidth]{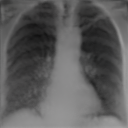}
  \caption{
    Kernel Inception Distance values with correlated example image generated by GAN trained on the small dataset.
    ({\bf a})~KID $\thickapprox$ 19,462;
    ({\bf c})~KID $\thickapprox$ 13,294;
    ({\bf b})~KID $\thickapprox$ 12,262.
  }
  \label{fig:fid_progress_1000}
\end{figure}

\begin{figure}[tb]
  \centering
  
  \setlength{\unitlength}{1mm}
  {\bf a}\hspace{0.3em} \includegraphics[width=0.2\textwidth]{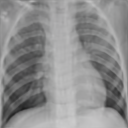} \hspace{1em}
  {\bf b}\hspace{0.3em} \includegraphics[width=0.2\textwidth]{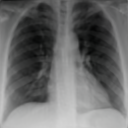} \hspace{1em}
  {\bf c}\hspace{0.3em} \includegraphics[width=0.2\textwidth]{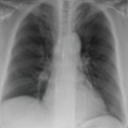}
  \caption{
    Kernel Inception Distance values with correlated example image generated by GAN trained on the micro dataset.
    ({\bf a})~KID $\thickapprox$ 18,826;
    ({\bf b})~KID $\thickapprox$ 13,299;
    ({\bf c})~KID $\thickapprox$ 12,89.
  }
  \label{fig:fid_progress_500}
\end{figure}

After the training had been finished we generated 1000 artificial images per class for both experiments, with example images presented in Fig.~\ref{fig:fid_progress_1000} and ~\ref{fig:fid_progress_500}. The GAN-augmented training dataset for CNN contained 2000 images per class (8000 images in total) for the small dataset and 1500 images per class (6000 in total) for the micro dataset. 

\subsection{Image comparison metrics}

\begin{figure}[tb] % avoid the location qualifier "h" unless strictly necessary
  \centering
  \includegraphics[width=0.7\textwidth]{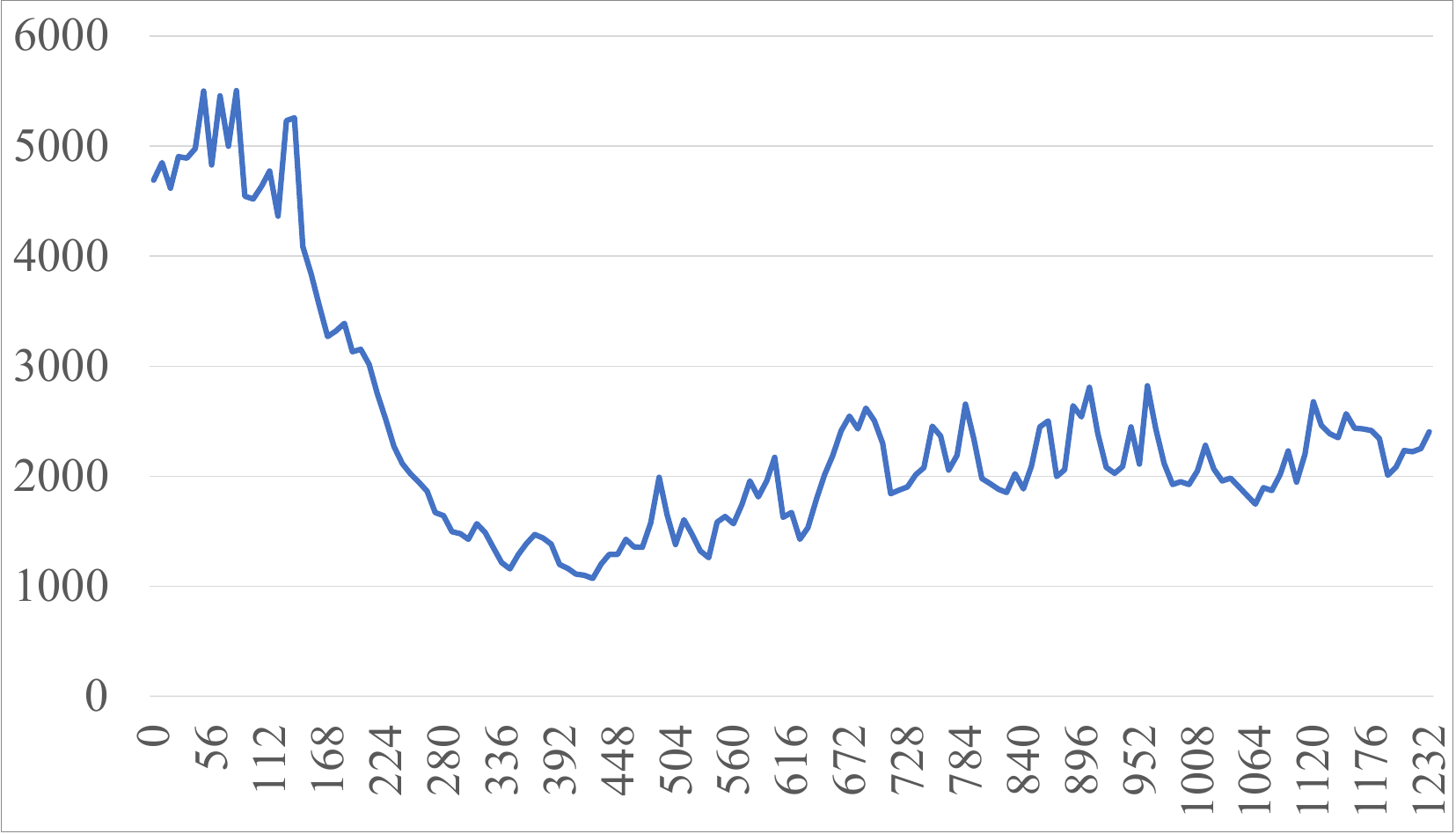} % recommended image formats are eps and pdf without lossless compression; please provide figures in which no artefacts are seen
  \caption{Fréchet Inception Distance graph for the small dataset (1000 images per class in the train subset) GAN training}
  \label{fig:single_fid_1000}
\end{figure}

\begin{figure}[tb] % avoid the location qualifier "h" unless strictly necessary
  \centering
  \includegraphics[width=0.7\textwidth]{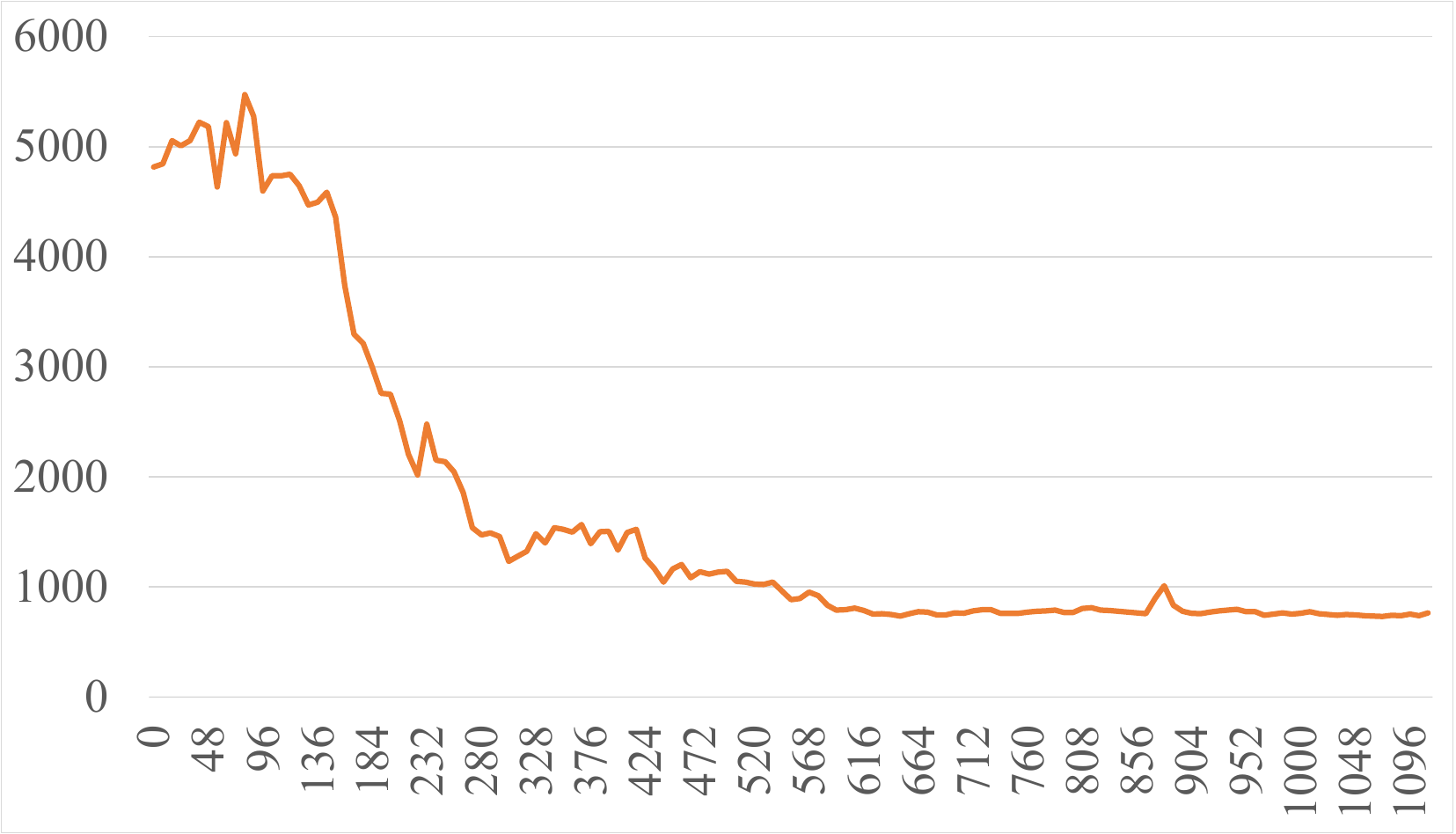} % recommended image formats are eps and pdf without lossless compression; please provide figures in which no artefacts are seen
  \caption{Fréchet Inception Distance graph for the micro dataset(500 images per class in the train subset) GAN training}
  \label{fig:single_fid_500}
\end{figure}

\begin{figure}[tb] % avoid the location qualifier "h" unless strictly necessary
  \centering
  \includegraphics[width=0.7\textwidth]{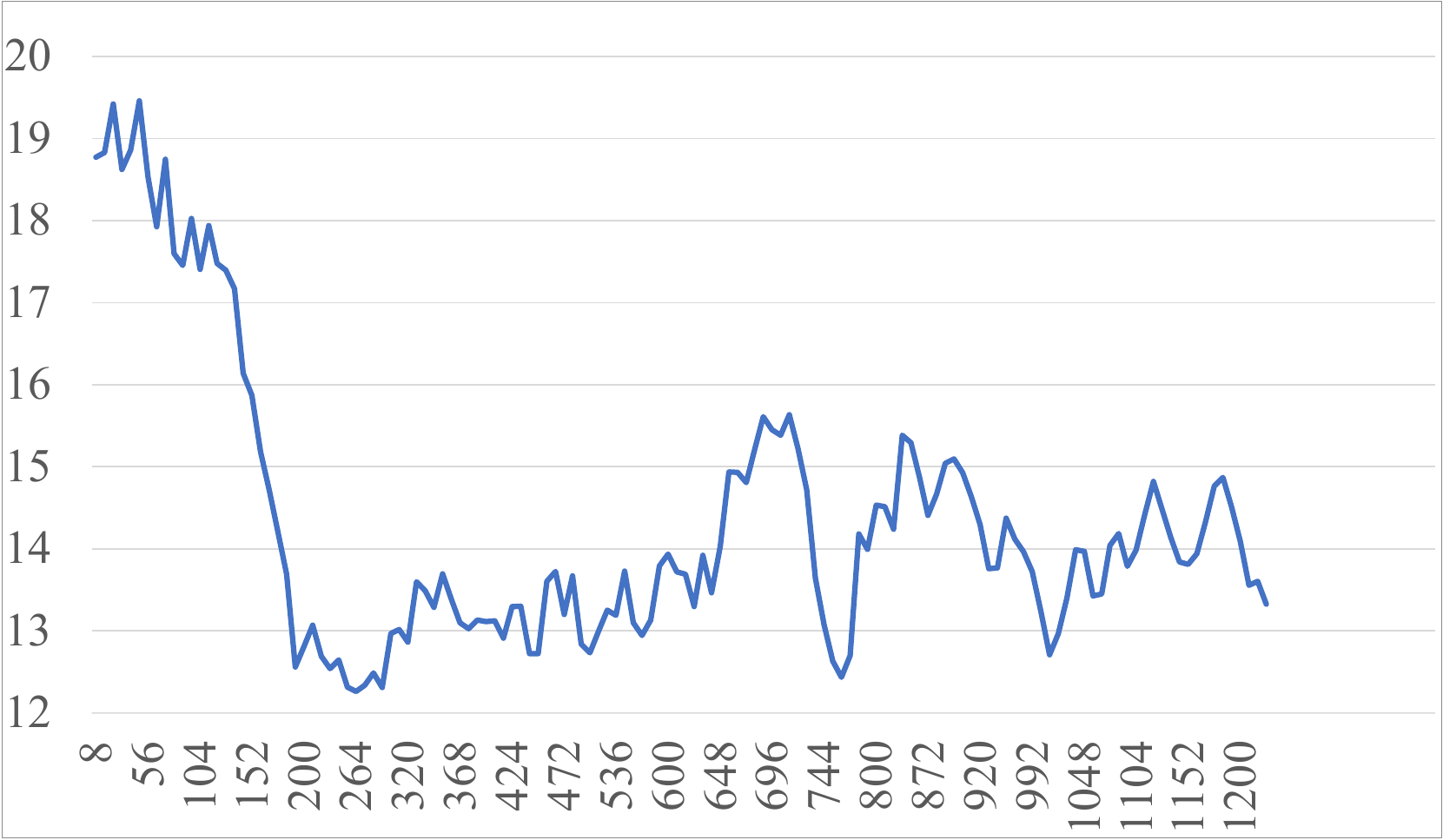} % recommended image formats are eps and pdf without lossless compression; please provide figures in which no artefacts are seen
  \caption{Kernel Inception Distance graph for case the small dataset (1000 images per class in the train subset) GAN training}
  \label{fig:single_kid_1000}
\end{figure}

\begin{figure}[tb] % avoid the location qualifier "h" unless strictly necessary
  \centering
  \includegraphics[width=0.7\textwidth]{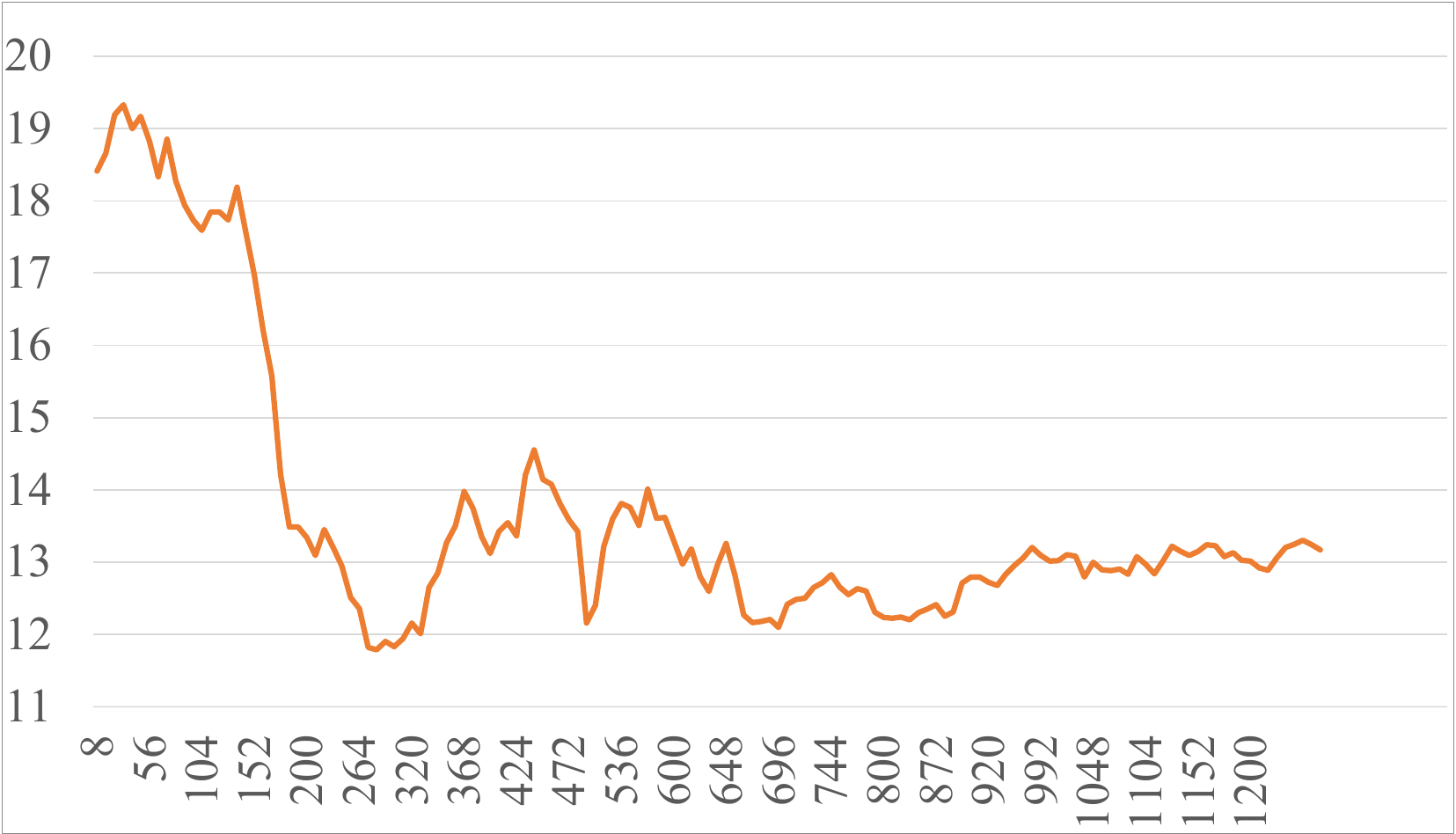} % recommended image formats are eps and pdf without lossless compression; please provide figures in which no artefacts are seen
  \caption{Kernel Inception Distance graph for the micro dataset (500 images per class in the train subset) GAN training}
  \label{fig:single_kid_500}
\end{figure}

In the previous work, we have used the Fréchet Inception Distance (FID)~\cite{heusel2018gansFID} metric to select the best-performing state of the StyleGAN2-ADA network~\cite{FirstPaper} throughout the training process as it is commonly used to evaluate the quality of images generated by GANs. We calculated a mean FID value, for each training epoch, across all classes between the train subset and generated images.
In this research, we added the Kernel Inception Distance (KID)~\cite{binkowski2021demystifyingKID} metric as FID is biased for smaller datasets~\cite{chong2020effectivelyFIDBias}. KID is very similar to FID in that it measures the difference between two sets of samples by calculating the square of the maximum mean discrepancy between vectors of vision-relevant features as extracted by the Inception-v3~\cite{InceptionV3} classifier network. KID compared to FID has several advantages and performs better with smaller sets and more consistently matches human perception. Similarly to FID, a smaller value of KID means that compared images are more similar to each other, and comparing the same set of images will result in a value equal to $0$.
Looking at the graphs of both KID and FID values per epoch (Figures~\ref{fig:single_fid_1000}-\ref{fig:single_kid_500}) it is visible that the overall training trend is the same - the value drops with each epoch of training until around epoch 250 and then it starts to grow slowly. But at the same time, KID values change more drastically per each epoch which may indicate, that the KID metric is more sensitive to differences between real and generated images. In addition, as in the original paper, we used RMSE, SRE, and SSIM metrics to verify the quality of generated images ~\cite{FirstPaper}.

% KERNEL INCEPTION DISTANCE

\subsection{Classical augmentation}

\begin{figure}[tb]
  \centering
  
  \setlength{\unitlength}{1mm}
  \hspace{0.3em} \includegraphics[width=0.2\textwidth]{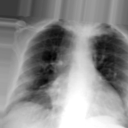} \hspace{1em}
  \hspace{0.3em} \includegraphics[width=0.2\textwidth]{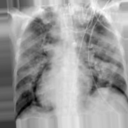} \hspace{1em}
  \hspace{0.3em} \includegraphics[width=0.2\textwidth]{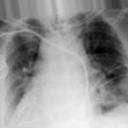}
  \caption{
    Examples of images with applied classical augmentation pipeline. 
  }
  \label{fig:classical_augmentation_examples}
\end{figure}

Similarly to the GAN-based augmentation described earlier, we've generated 1000 additional images (Fig.~\ref{fig:classical_augmentation_examples}) by applying classical augmentation with parameters as follows: 

\begin{itemize}
    \item rotation - randomly rotate the image by an angle of up to 5 degrees clockwise or counterclockwise;
    \item shift - randomly shift the image along cardinal axes within the range of 5$\%$ of the specific image size, the empty field is filled with the trace of the last shifted pixels;
    \item stretch - randomly stretch the image between opposite vertices by up to 5 $\%$;
    \item zoom - randomly zoom inside pictures up to 15$\%$ of the specific value of the image size;
    \item brightness change - randomly enlighten or darken the image by up to $40\%$.
\end{itemize}

The values of the parameters remain the same as in the previous work and were picked to maximize the accuracy score on the validation subset~\cite{FirstPaper}.

\subsection{Classification evaluation}

To evaluate and compare the augmentation techniques described above, we trained a convolutional neural network using each of them. The network was trained to classify 4 classes present in the original dataset. We used Inception-v3 with a transfer learning technique as an augmentation benchmark network~\cite{InceptionV3}. We used Keras default implementation with library-provided ImageNet weights~\cite{InceptionV3Keras}. The head of the network was replaced with the following output layers for the transfer learning: 

\begin{itemize}
    \item Flatten layer
    \item Dense layer with 10 neurons and an Exponential Linear Unit (ELU) activation function~\cite{ELU}
    \item Dense layer with 20 neurons and an ELU activation function
    \item Dense layer with 4 neurons and a softmax activation function
\end{itemize}

Categorical cross-entropy was picked as the loss function and the learning rate was set to Keras default of 0.001 ~\cite{RMSpropKeras}.
We continue using RMSprop as an optimizer as it was selected on validation accuracy in the original article ~\cite{FirstPaper}. The network was trained for 11 epochs with a batch size equal to 32. The total amount of epochs was reduced in comparison with previous work as the best results were achieved in the first 11 epochs in almost all training runs. For the clarity of the experiment, the network was trained 5 times for each experimental case and each augmentation pipeline (no augmentation, classical augmentation, GAN-augmentation). The network frozen weights with the highest validation accuracy score were picked as a final version that was later evaluated on the test subset. To examine and compare the quality of the trained network, several model evaluation metrics were calculated: accuracy, precision, recall, F1, specificity, and Matthew's correlation coefficient (MCC). Metrics were calculated on the test dataset. Values of the calculated metrics are presented in the Results section of the paper. 

\section{Results}

\begin{figure}
  \setlength{\unitlength}{1mm}
  {\bf a}\hspace{0.3em} \includegraphics[width=0.45\textwidth]{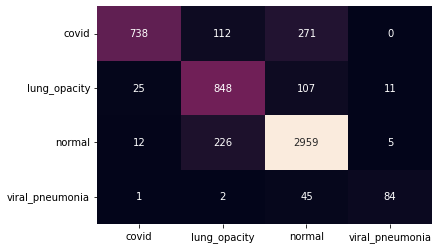} \hspace{1em}
  {\bf b}\hspace{0.3em} \includegraphics[width=0.45\textwidth]{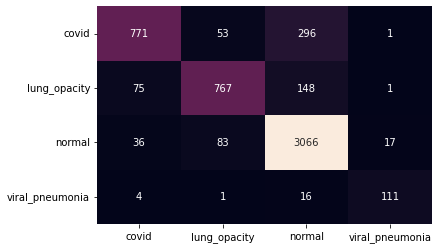} \hspace{1em}

  \centering
  {\bf c}\hspace{0.3em} \includegraphics[width=0.45\textwidth]{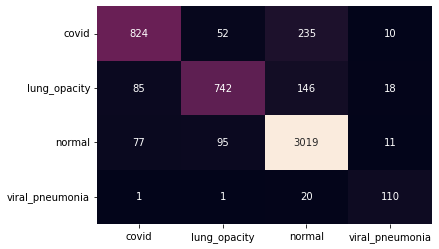} \hspace{1em}
  \caption{
    Confusion matrices for different augmentations of the small dataset. The vertical axis represents predicted values, horizontal axis represents real values. 
    ({\bf a})~ No augmentations;
    ({\bf b})~ Classical augmentations;
    ({\bf c})~ GAN-based augmentations. 
  }
  \label{fig:conf_matrix_small}
\end{figure}

\begin{figure}
  \setlength{\unitlength}{1mm}
  {\bf a}\hspace{0.3em} \includegraphics[width=0.45\textwidth]{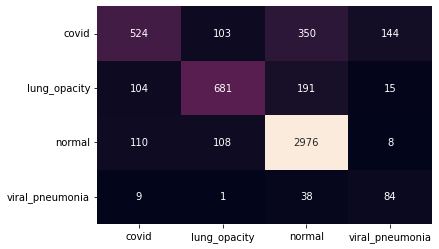} \hspace{1em}
  {\bf b}\hspace{0.3em} \includegraphics[width=0.45\textwidth]{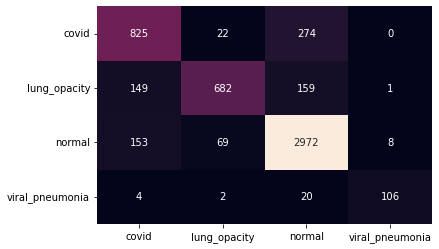} \hspace{1em}

  \centering
  {\bf c}\hspace{0.3em} \includegraphics[width=0.45\textwidth]{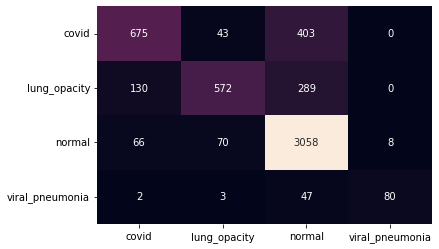} \hspace{1em}
  \caption{
    Confusion matrices for different augmentations of the micro dataset. The vertical axis represents predicted values, horizontal axis represents real values. 
    ({\bf a})~ No augmentations;
    ({\bf b})~ Classical augmentations;
    ({\bf c})~ GAN-based augmentations. 
  }
  \label{fig:conf_matrix_micro}
\end{figure}

After the classification metrics are calculated on the test dataset, it is visible, that any augmentation approach is better than no augmentation at all. At the same time, GAN-based augmentation and classical augmentation perform comparably for the dataset with at least 1000 images per class in the training dataset, as shown in Table~\ref{tab:results_small} and Fig.~\ref{fig:conf_matrix_small}. Classical augmentation outperforms GAN-based augmentation with datasets containing 500 images per class, as presented in Table~\ref{tab:results_micro} and Fig.~\ref{fig:conf_matrix_micro}. The overall networks' results are slightly worse independently of the augmentation approach in comparison with the ones achieved in our previous work ~\cite{FirstPaper}. Finally, the classical augmentation shows itself as the best augmentation approach while GAN-based augmentation can achieve comparable results but requires significantly more time and hardware to be performed.

\begin{table}[h] % avoid the location qualifier "h" unless strictly necessary
  \centering
  \caption{Classification metrics values for the small dataset (1000 images in the train subset)}
  \label{tab:results_small}
  \begin{tabular}{lllllll}
    \hline\hline
    Augmentation pipeline  & Accuracy & Precision & Recall & F1    & Specificity & MCC \\ \hline
    No augmentation        & 0,85     & 0,845     & 0,769  & 0,805 & 0,931       & 0,74 \\
    Classical augmentation & 0,87     & 0,861     & 0,815  & 0,837 & 0,934       & 0,78 \\
    GAN-augmentation       & 0,862    & 0,822     & 0,815  & 0,819 & 0,936       & 0,76 \\
    \hline\hline
  \end{tabular}
\end{table}

\begin{table}[h] % avoid the location qualifier "h" unless strictly necessary
  \centering
  \caption{Classification metrics values for the micro dataset (500 images in the train subset)}
  \label{tab:results_micro}
  \begin{tabular}{lllllll}
    \hline\hline
    Augmentation pipeline  & Accuracy  & Precision   & Recall   & F1      & Specificity & MCC  \\ \hline
    No augmentation        & 0,783     & 0,659       & 0,68     & 0,669   & 0,903       & 0,573 \\
    Classical augmentation & 0,842     & 0,85        & 0,789    & 0,818   & 0,93        & 0,749 \\
    GAN-augmentation       & 0,81      & 0,83        & 0,685    & 0,751   & 0,9         & 0,665 \\
    \hline\hline
  \end{tabular}
\end{table}

We can take a look (Table~\ref{tab:results_original}) at the accuracy value obtained in our previous work where 2000 images per class were used ~\cite{FirstPaper}, there is a visible correlation between classification accuracy and the size of the training dataset independently from data augmentation applied.

\begin{table}[h]
\centering
  \caption{Accuracy metric value from the original article (2000 images per dataset)}
  \label{tab:results_original}
\begin{tabular}{lllllll}
\hline\hline
                Augmentation pipeline  & Accuracy \\ %  & precision   & recall      & f1          & specificity & mcc \\
                No augmentation      & $0.855$ \\ %& $0.859$     & $0.860$     & $0.860$     & $0.951$     & $0.810$ \\
                Classic augmentation & $0.891$ \\ % & $0.896$     & $0.893$     & $0.895$     & $0.963$     & $0.858$ \\
                GAN-augmentation     & $0.871$ \\ %& $0.886$     & $0.874$     & $0.880$     & $0.956$     & $0.836$ \\
\hline\hline

\end{tabular}
\end{table}

\section{Conclusion}
\label{sec:conclusion}

We've studied the performance of GAN-based augmentation for the classification of lung X-ray medical images as a function of dataset size. The obtained results show that GAN-based augmentation is comparable with classical augmentation for medium and large datasets. Unfortunately, the time and hardware requirements make it unreasonable to use such an approach as the main augmentation technique. In the case of small datasets, the GAN model wasn't able to train well enough to be a source of valuable training data. At the same time, the fact of GAN being able to compete with classical augmentation for larger datasets, potentially allows researchers and medical institutions to solve the problem of medical data availability by sharing synthetically generated images instead of real ones~\cite{ganPrivacy}. Therefore, the topic of GAN-based augmentation should be investigated further.

\subsubsection*{Acknowledgment}

This work was completed in part with resources provided by the Świerk Computing Centre at the National Centre for Nuclear Research.
This work benefited from the software tools developed in the frame of the EuroHPC PL Project, Smart Growth Operational Programme 4.2.
We gratefully acknowledge Polish high-performance computing infrastructure PLGrid (HPC Centers: ACK Cyfronet AGH) for providing computer facilities and support within computational grant no. PLG/2022/015617.
  
\bibliographystyle{habbrv}
\bibliography{main}

\end{document}